\documentstyle[12pt,twoside,fleqn,epsf,espcrc1]{article}

\def\gsim{\mathrel{\raise.3ex\hbox{$>$}\mkern-14mu
             \lower0.6ex\hbox{$\sim$}}}
\def\lsim{\mathrel{\raise.3ex\hbox{$<$}\mkern-14mu
             \lower0.6ex\hbox{$\sim$}}}

\newcommand{\AmS}{{\protect\the\textfont2
  A\kern-.1667em\lower.5ex\hbox{M}\kern-.125emS}}

\newcommand{\ga}{\gamma}
\newcommand{\bk}{{\bf k}}

\newcommand{\psib}{{\bar\psi}}

\renewcommand{\slash}{\!\!\!\!/\,}

\newcommand{\dsp}{\displaystyle}

\title{The Recent Excitement in High-Density QCD
\thanks{Invited talk at PANIC
`99, Uppsala, Sweden, June 1999. IASSNS-HEP/99-68}}

\author{Frank Wilczek\address{School of Natural Science, 
        Institute for Advanced Study\\ 
        Princeton, NJ 08540 USA}
        \thanks{Research supported in part by DOE grant
                DE-FG02-90ER40542. e-mail: wilczek@sns.ias.edu}}
       
\begin{document}

\maketitle

\begin{abstract}

Over the past few months, the theory of QCD at high density
has been advanced considerably.  It provides new perspectives on, and
controlled realizations of, confinement and chiral symmetry breaking.
Here I survey the recent developments, and suggest a few directions
for future work.

\end{abstract}

\section{Introduction}

The behavior of QCD at high density is intrinsically interesting, as
the answer to the question: What happens to matter, if you keep
squeezing it harder and harder?  It is also directly relevant to the
description of neutron star interiors, neutron star collisions, and
events near the core of collapsing stars.  Also, one might hope to
obtain some insight into physics at ``low'' density -- that is,
ordinary nuclear density or just above -- by approaching it from the
high-density side.

Why might we anticipate QCD simplifies in the limit of high density?
A crude answer is: ``Asymptotic freedom meets the fermi
surface.''   One might argue, formally, that the only external
mass scale characterizing the problem is the large chemical potential
$\mu$, so that if the effective coupling $\alpha_s(\mu)$ is small, as
it will be for $\mu \gg \Lambda_{QCD}$, where $\Lambda_{QCD} \approx
200$ Mev is the primary QCD scale, then we have a weak coupling
problem.  More physically, one might argue that at large $\mu$ the
relevant, low-energy degrees of freedom involve modes near the fermi
surface, which have large energy and momentum.  An interaction between
particles in these modes will either barely deflect them, or will
involve a large momentum transfer.  In the first case we don't
care, while the second is governed by a small effective coupling.

These arguments are too quick, however.  The formal argument is
specious, if the perturbative expansion contains infrared divergences.
And there are good reasons -- two separate ones, in fact -- to
anticipate such divergences.

First, fermi balls are generically unstable against the effect of
attractive interactions, however weak, between pairs near the fermi
surface that carry equal and opposite momentum.  This is the Cooper
instability, which drives ordinary superconductivity in metals and the
superfluidity of He3.  It is possible because occupied pair states can
have very low energy, and they can all scatter into one another.  Thus
one is doing highly degenerate perturbation theory, and in such a
situation even a very weak coupling can produce significant
``nonperturbative'' effects.

Second, nothing in our heuristic argument touches the gluons.  To be
sure the gluons will be subject to electric screening, but at zero
frequency there is no magnetic screening, and infrared divergences do
in fact arise, through exchange of soft magnetic gluons.

Fortunately, by persisting along this line of thought we find a path
through the apparent difficulties.  Several decades ago Bardeen,
Cooper, and Schrieffer taught us, in the context of metallic
superconductors, how the Cooper instability is resolved
\cite{schrieffer}. We can easily adapt their methods to QCD
\cite{frautschi} \cite{bailin}.  In electronic systems only rather
subtle mechanisms can generate an attractive effective interaction
near the fermi surface, since the primary electron-electron
interaction is Coulomb repulsion.  In QCD, remarkably, it is much more
straightforward.  Even at the crudest level we find attraction.
Indeed, two quarks, each a color triplet, can combine to form a single
color antitriplet, thus reducing their total field energy.

The true ground state of the quarks is quite different from the naive
fermi balls.  It is characterized by the formation of a coherent
condensate, and the development of an energy gap.  The condensation,
which is energetically favorable, is inconsistent with a magnetic
field, and so weak magnetic fields are expelled.  This is the famous
Meissner effect in superconductivity, which is essentially identical
to what is known as the Higgs phenomenon in particle physics.
Magnetic screening of gluons, together with energy gaps for quark
excitations, remove the potential sources of infrared divergences
mentioned above.  Thus we have good reasons to hope that a weak
coupling -- though, of course, nonperturbative -- treatment of the
high density state will be fully consistent and accurate.

The central result in the recent developments is that this program can
be carried to completion rigorously in QCD with sufficiently many
(three or more) quark species \cite{cfl}.  Thus the more refined, and
fully adequate, answer to our earlier question is: ``Asymptotic
freedom meets the BCS groundstate.''  Together, these concepts
render the behavior of QCD at asymptotically high density calculable.

The simplest and most beautiful results, luckily, occur in the version
of QCD containing three quarks having equal masses.  I say luckily,
because this idealization applies to the real world, at densities so
high that we can neglect the strange quark mass (yet not so high that
we have to worry about charmed quarks).  Here we encounter the
phenomenon of color-flavor locking.  The ground state contains
correlations whereby both color and flavor symmetry are spontaneously
broken, but the diagonal subgroup, which applies both transformations
simultaneously, remains valid.

Color-flavor locking has many remarkable consequences \cite{cfl}.
There is a gap for all colored excitations, including the gluons.
This is, operationally, confinement.  The photon picks up a gluonic
component of just such a form as to ensure that all elementary
excitations, including quarks, are integrally charged.  Some of the
gluons acquire non-zero, but integer-valued, electric charges.  Baryon
number is spontaneously broken, which renders the high-density
material a superfluid.

If in addition the quarks are massless, then their chiral symmetry is
spontaneously broken, by a new mechanism.  The left-dynamics and the
right-dynamics separately lock to color; but since color allows only
vector transformations, left is thereby locked to right.

You may notice several points of resemblance between the low-energy
properties calculated for the high-density color-flavor locked phase
and the ones you might expect at low density, based on
semi-phenomenological considerations such as the MIT bag model, or
experimental results in real-world QCD.  The quarks play the role of
low-lying baryons, the gluons play the role of the low-lying vector
mesons, and the Nambu-Goldstone bosons of broken chiral symmetry play
the role of the pseudoscalar octet.  All the quantum numbers match,
and the spectrum has gaps -- or not -- in all the right places.  In
addition we have baryon number superfluidity, which extends the
expected pairing phenomena in nuclei.  Overall, there is an uncanny
match between all the universal, and several of the non-universal,
features of the calculable high-density and the expected low-density
phase.  This leads us to suspect that there is no phase transition
between them \cite{cont}!

\section{Sculpting the Problem}
\subsection{renormalization group toward the fermi surface}
To sculpt the problem, begin by assuming weak coupling, and focus on
the quarks.  Then the starting point is fermi balls for all the
quarks, and the low-energy excitations include states where some modes
below the nominal fermi surface are vacant and some modes above are
occupied.  The renormalization group, in a generalized sense, is a
philosophy for dealing with problems involving nearly degenerate
perturbation theory.  In this approach, one attempts to map the
original problem onto a problem with fewer degrees of freedom, by
integrating out the effect of the higher-energy (or, in a relativistic
theory, more virtual) modes.  Then one finds a new formulation of the
problem, in a smaller space, with new couplings.  In favorable cases
the reformulated problem is simpler than the original, and one can go
ahead and solve it.

This account of the renormalization group might seem odd, at first
sight, to high-energy physicists accustomed to using asymptotic
freedom in QCD.  That is because in traditional perturbative QCD one
runs the procedure backward.  When one integrates out highly virtual
modes, one finds the theory becomes more strongly coupled.  Simplicity
arises when one asks questions that are somehow inclusive, so that to
answer them one need not integrate out very much.  It is then that the
microscopic theory, which is ideally symmetric and constrained,
applies directly.  So one might say that the usual application of the
renormalization group in QCD is fundamentally negative: it informs us
how the fundamentally simple theory comes to look complicated at low
energy, and helps us to identify situations where we can avoid the
complexity.

Here, although we are still dealing with QCD, we are invoking quite a
different renormalization group, one which conforms more closely to
the Wilsonian paradigm \cite{shankar} \cite{polchinski}.  We consider
the effect of integrating out modes whose energy is within the band
$(\epsilon, \delta \epsilon)$ of the fermi surface, on the modes of
lower energy.  This will renormalize the couplings of the remaining
modes, due to graphs like those displayed in Figure 1.  In addition
the effect of higher-point interactions is suppressed, because the
phase space for them shrinks, and it turns out that only four-fermion
couplings survive unscathed (they are the marginal, as opposed to
irrelevant, interactions).  Indeed the most significant interactions
are those involving particles or holes with equal and opposite three
momenta, since they can scatter through many intermediate states.  For
couplings $g_\eta$ of this kind we find
\begin{equation}
\label{fermiRG}
{d g_\eta \over d \ln \delta } = \kappa_\eta g_\eta^2. 
\end{equation}
Here $\eta$ labels the color, flavor, angular momentum, ...  channel
and in general we have a matrix equation \cite{evans} \cite{swRG} --
but let's keep it simple, so $\kappa_\eta$ is a positive number.
Then \ref{fermiRG} is quite simple to integrate, and we have
\begin{equation}
{1\over g_\eta (1) } - {1\over g_\eta(\delta ) } =  \kappa_\eta \ln \delta.
\end{equation}
Thus for $g_\eta(1)$ negative, corresponding to attraction, $|g_\eta
(\delta)|$ will grow as $\delta \rightarrow 0$, and become singular
when
\begin{equation}
\delta =  e^{1\over \kappa_\eta g_\eta (1)}.
\end{equation}
Note that although the singularity occurs for arbitrarily weak
attractive coupling, it is nonperturbative.

\subsection{model hamiltonian and condensation}
The renormalization group toward the fermi surface helps us identify
potential instabilities, but it does not indicate how they are
resolved.  The great achievement of BCS was to identify the form of
the stable ground state the Cooper instability leads to.  Their
original calculation was variational, and that is still the most
profound and informative approach, but simpler, operationally
equivalent algorithms are now more commonly used.  I will be very
sketchy here, since this is textbook material.

Most calculations to date have been based on model interaction
Hamiltonians, that are motivated, but not strictly derived, from
microscopic QCD.  They are chosen as a compromise between realism and
tractability.  For concreteness I shall here follow \cite{cfl}, and
consider
\begin{equation}
\label{modelH}
\begin{array}{ll}
H &= \dsp\int d^3x\, \psib(x) (
\nabla\slash - \mu\ga_0) \psi(x) + H_I, \\
H _I &= K \dsp\sum_{\mu, A}\int d^3x\, {{\cal F}}
 \psib(x) \ga_\mu T^A \psi(x)\ \psib(x) \ga^\mu T^A \psi(x)\ 
\end{array}
\end{equation}
Here the $T^A$ are the color $SU(3)$ generators, so the quantum
numbers are those of one-gluon exchange.  However instead of an honest
gluon propagator we use an instantaneous contact interaction, modified
by a form-factor $\cal F$.  $\cal F$ is taken to be a product of
several momentum dependent factors $F(p)$, one for each leg, and to
die off at large momentum.  One convenient possibility is $F(p) =
(\lambda^2/(p^2 + \lambda^2))^\nu$, where $\lambda$ and $\nu$ can be
varied to study sensitivity to the location and shape of the cutoff.
The qualitative effect of the form-factor is to damp the spurious
ultraviolet singularities introduced by $H_I$; microscopic QCD, of
course, does have good ultraviolet behavior.  One will tend to trust
conclusions that do not depend sensitively on $\lambda$ or $\nu$. In
practice, one finds that the crucial results -- the form and magnitude
of gaps -- are rather forgiving.

Given the Hamiltonian, we can study the possibilities for symmetry
breaking condensations.  The most favorable condensation possibility
so far identified is of the form
\begin{equation}
\label{condensate}
\begin{array}{lll}
~\langle q^{i\alpha}_{La} (p) q^{j\beta}_{Lb} (-p) \rangle ~=~ 
-\langle q^{i\alpha}_{Ra} (p) q^{j\beta}_{Rb} (-p) \rangle ~=~ 
\epsilon^{ij} (\kappa_1(p^2) \delta^\alpha_a \delta^\beta_b +
\kappa_2(p^2) \delta^\alpha_b \delta^\beta_a)~. 
\end{array}
\end{equation}
There are several good reasons to think that condensation of this form
characterizes the true ground state, with lowest energy, at asymptotic
densities.  It corresponds to the most singular channel, in the
renormalization group analysis discussed above.  It produces a gap in
all channels, and is perturbatively stable, so that it is certainly a
convincing local minimum.  And it beats various more-or-less plausible
competitors that have been investigated, by a wide margin.

Given the form of the condensate, one can fix the leading functional
dependencies of $\kappa_1(p^2, \mu)$ and $\kappa_2(p^2, \mu)$ at weak
coupling by a variational calculation.  For present purposes, it is
adequate to replace all possible contractions of the quark fields in
\ref{modelH} having the quantum numbers of \ref{condensate} with their
supposed expectation values, and diagonalize the quadratic part of the
resulting Hamiltonian.  The ground state is obtained, of course, by
filling the lowest energy modes, up to the desired density.  One then
demands internal consistency, {\it i.e}. that the postulated
expectation values are equal to the derived ones.  Some tricky but
basically straightforward algebra leads us to the result
\begin{equation}
\label{pDependence}
\Delta_{1,8} (p^2) = F(p)^2 \Delta_{1,8} 
\end{equation}
where $\Delta_1$ and $\Delta_8$ satisfy the coupled gap equations
\begin{equation}
\begin{array}{ll}
 \Delta_8 + {1\over 4} \Delta_1 &  = {16\over 3}K G(\Delta_1) \\ 
\noalign{\medskip}
{1\over 8} \Delta_1  &  =  {16\over 3}K G(\Delta_8) 
\end{array}
\end{equation}
where we have defined
\begin{equation}
\label{coupGaps}
G(\Delta) = -{1\over 2}\sum_{\bk}\Biggl\{
  {F(k)^4 \Delta\over \sqrt{ (k-\mu)^2 + F(k)^4 \Delta^2}}  
+ {F(k)^4 \Delta\over \sqrt{ (k+\mu)^2 + F(k)^4 \Delta^2} }\Biggr\}
\end{equation}
and
\begin{equation}
\begin{array}{ll}
\kappa_1(p^2) & = {1\over 8K}(\Delta_8(p^2) + {1\over 8} \Delta_1(p^2)) \\
\noalign{\medskip}
\kappa_2(p^2) & = {3\over 64K} \Delta_1(p^2).
\end{array}
\end{equation}
The $\Delta$ are defined so that $F(p)^2 \Delta_{1,8}(p^2)$ are the
gaps for singlet or octet excitations at 3-momentum $p$.   
Equations \ref{coupGaps}  must be solved numerically.  

Finally, to obtain quantitative estimates of the gaps, we must
normalize the parameters of our model Hamiltonian.  One can do this
very crudely by using the model Hamiltonian in the manner originally
pioneered by Nambu and Jona-Lasinio \cite{nambu}, that is as the basis
for a variational calculation of chiral symmetry breaking at zero
density.  The magnitude of this chiral condensate can then be fixed to
experimental or numerical results.  In this application we have no
firm connection between the model and microscopic QCD, because there
is no large momentum scale (or weak coupling parameter) in sight.
Nevertheless a very large literature following this approach
encourages us to hope that its results are not wildly wrong,
quantitatively.  Upon adopting this normalization procedure, one finds
that gaps of order several tens of Mev near the fermi surface are
possible at moderate densities.

While this model treatment captures major features of the physics of
color-flavor locking, with a little more work it is possible to do a
much more rigorous calculation, and in particular to normalize
directly to the known running of the coupling at large momentum.  This
will be sketched below.

\section{Consequences of Color-Flavor Locking: Symmetry}
\subsection{broken gauge invariance?}

An aspect of \ref{condensate} that might appear troubling at first
sight, is its lack of gauge invariance.  There are powerful general
arguments that local gauge invariance cannot be broken \cite{elitzur}.
Indeed, local gauge invariance is really a tautology, stating the
equality between redundant variables.  Yet its `breaking' is
central to two of the most successful theories in physics, to wit BCS
superconductivity theory and the standard model of electroweak
interactions.  In BCS theory we postulate a non-zero vacuum
expectation value for the (electrically charged) Cooper pair field,
and in the standard model we postulate a non-zero vacuum expectation
value for the Higgs field, which violates both the weak isospin SU(2)
and the weak hypercharge U(1).

In each case, we should interpret the condensate as follows.  We are
working in a gauge theory at weak coupling.  It is then very
convenient to fix a gauge, because after we have done so -- but not
before! -- the gauge potentials will make only small fluctuations
around zero, which we will be able to take into account
perturbatively.  Of course at the end of any calculation we must
restore the gauge symmetry, by averaging over the gauge fixing
parameters (gauge unfixing).  Only gauge-invariant results will
survive this averaging.  In a fixed gauge, however, one might capture
important correlations, that characterize the ground state, by
specifying the existence of non-zero condensates relative to that
gauge choice.  These condensates need not, and generally will not,
break any symmetries.

For example, in the standard electroweak model one employs a non-zero
vacuum expectation value for a Higgs doublet field $\langle \phi^a
\rangle ~=~ v \delta^a_1$, which is not gauge invariant.  One might be
tempted to use the magnitude of its absolute square, which is gauge
invariant, as an order parameter for the symmetry breaking, but
$\langle \phi ^\dagger \phi \rangle $ never vanishes, whether or not
any symmetry is broken (and, of course, $\langle \phi ^\dagger \phi
\rangle $ breaks no symmetry).  In fact there is no order parameter
for the electroweak phase transition, and it has long been appreciated
\cite{fradkin} that one could, by allowing the $SU(2)$ gauge couplings
to become large, go over into a `confined' regime while
encountering no sharp phase transition.  The most important
gauge-invariant consequences one ordinarily infers from the
condensate, of course, are the non-vanishing W and Z boson masses.
This absence of massless bosons and long-range forces is the essence
of confinement, or of the Meissner-Higgs effect.  Evidently, when used
with care, the notion of spontaneous gauge symmetry breaking can be an
extremely convenient fiction -- so it proves for \ref{condensate}.

\subsection{broken symmetries and consequences}

The equations of our original model, QCD with three massless flavors,
has the continuous symmetry group $SU(3)^c \times SU(3)_L \times
SU(3)_R \times U(1)_B$.  The Kronecker deltas that appear in the
condensate \ref{condensate} are invariant under neither color nor
left-handed flavor nor right-handed flavor rotations separately. Only
a global, diagonal $SU(3)$ leaves the ground state invariant. Thus we
have the symmetry breaking pattern
\begin{equation}
SU(3)^c \times SU(3)_L \times SU(3)_R \times U(1)_B \rightarrow
SU(3)_{c+L+R} \times Z_2~. 
\end{equation}

Indeed, if we make a left-handed chiral rotation we can compensate it
by a color rotation, to leave the left-handed condensate invariant.
Color rotations being vectorial, we must then in addition make a
right-handed chiral rotation, in order to leave the right-handed
condensate invariant.  Thus chiral symmetry is spontaneously broken,
by a new mechanism: although the left- and right- condensates are
quite separate (and, before we include instantons -- see below -- not
even phase coherent), because both are locked to color they are
thereby locked to one another.

The breaking of local color symmetry implies that all the gluons
acquire mass, according to the Meissner (or alternatively Higgs)
effect.  There are no long-range,$1/r$ interactions.  There is no
direct signature for the color degree of freedom -- although, of
course, in weak coupling one clearly perceives its avatars.  It is
veiled or, if you like, confined.

The spontaneous breaking of global chiral $SU(3)_L \times SU(3)_R$
brings with it an octet of pseudoscalar Nambu-Goldstone bosons,
collective modes interpolating, in space-time, among the condensates
related by the lost symmetry.  These massless modes, as is familiar,
are derivatively coupled, and therefore they do not generate singular
long-range interactions.

Less familiar, and perhaps disconcerting at first sight, is the loss
of baryon number symmetry.  This does not, however, portend proton
decay, any more than does the non-vanishing condensate of helium atoms
in superfluid ${\rm He4}$.  Given an isolated finite sample, the current
divergence equation can be integrated over a surface surrounding the
sample, and unambiguously indicates overall number conservation.  To
respect it, one should project onto states with a definite number of
baryons, by integrating over states with different values of the
condensate phase.  This does not substantially alter the physics of
the condensate, however, because the overlap between states of
different phase is very small for a macroscopic sample.  Roughly
speaking, there is a finite mismatch per unit volume, so the overlap
vanishes exponentially in the limit of infinite volume.  The true
meaning of the formal baryon number violation is that there are
low-energy states with different distributions of baryon number, and
easy transport among them.  Indeed, the dynamics of the condensate is
the dynamics of superfluidity:  gradients in the Nambu-Goldstone mode
are none other than the superfluid flow.

We know from experience that large nuclei exhibit strong even-odd
effects, and an extensive phenomenology has been built up around the
idea of pairing in nuclei.  If electromagnetic Coulomb forces
didn't spoil the fun, we could confidently expect that extended
nuclear matter would exhibit the classic signatures of superfluidity.
In our 3-flavor version the Coulomb forces do not come powerfully into
play, since the charges of the quarks average out to electric
neutrality.  Furthermore, the tendency to superfluidity exhibited by
ordinary nuclear matter should be enhanced by the additional channels
operating coherently.  So one should expect strong superfluidity at
ordinary nuclear density, and it becomes less surprising that we find
it at asymptotically large density too.
\subsection{true order parameters}

I mentioned before that the Higgs mechanism as it operates in the
electroweak sector of the standard model has no gauge-invariant
signature.  With color-flavor locking we're in better shape,
because global as well as gauge symmetries are broken.  Thus there are
sharp differences between the color-flavor locked phase and the free
phase. There must be phase transitions -- as a function, say, of
temperature -- separating them.

In fact, it is a simple matter to extract gauge invariant order
parameters from our primary, gauge variant condensate at weak
coupling.  For instance, to form a gauge invariant order parameter
capturing chiral symmetry breaking we may take the product of the
left-handed version of \ref{condensate} with the right-handed version
and saturate the color indices, to obtain
\begin{equation}
\label{chiral}
\langle q^\alpha_{La} q^\beta_{Lb} \bar q^c_{R\alpha} \bar q^d_{R\beta} \rangle
\sim \langle q^\alpha_{La} q^\beta_{Lb}\rangle \langle \bar
q^c_{R\alpha} \bar q^d_{R\beta} \rangle \sim (\kappa_1^2 + \kappa_2^2)
\delta^c_a \delta^d_b  
+ 2 \kappa_1 \kappa_2 \delta^d_a \delta^c_b
\end{equation}
Likewise we can take a product of three copies of the condensate and
saturate the color indices, to obtain a gauge invariant order
parameter for superfluidity.  These secondary order parameters will
survive gauge unfixing unscathed.  Unlike the primary condensate from
which they were derived, they are more than convenient fictions.
\subsection{a subtlety: axial baryon number}
As it stands the chiral order parameter \ref{chiral} is not quite the
usual one, but roughly speaking its square.  It leaves invariant an
additional $Z_2$, under which the left-handed quark fields change
sign.  Actually this $Z_2$ is not a legitimate symmetry of the full
theory, but suffers from an anomaly.

Since we can be working at weak coupling, we can be more specific.
Our model Hamiltonian \ref{modelH} was abstracted from one-gluon
exchange, which is the main interaction among high-energy quarks in
general, and so in particular for modes near our large fermi surfaces.
The instanton interaction is much less important, at least
asymptotically, both because it is intrinsically smaller for energetic
quarks, and because it involves six fermion fields, and hence (one can
show) is irrelevant as one renormalizes toward the fermi surface.
However, it represents the leading contribution to axial baryon number
violation.  In particular, it is only $U_A(1)$ violating interactions
that fix the relative phase of our left- and right- handed
condensates.  So a model Hamiltonian that neglects them will have an
additional symmetry that is not present in the full theory.  After
spontaneous breaking, which does occur in the axial baryon number
channel, there will be a Nambu-Goldstone boson in the model theory,
that in the full theory acquires an anomalously (pun intended)
small mass \cite{pisarski}.  Similarly, in the full theory there will
be a non-zero tertiary chiral condensate of the usual kind, bilinear
in quark fields, but it will be parametrically smaller than
\ref{chiral}.

\section{Consequences of Color-Flavor Locking: Elementary Excitations}

There are three sorts of elementary excitations.  They are the modes
produced directly by the fundamental quark and gluon fields, and the
collective modes connected with spontaneous symmetry breaking.

The quark fields of course produce spin $1/2$ fermions.  Some of these
are true long-lived quasiparticles, since there is nothing for them to
decay into.  They form an octet and a singlet under the residual
diagonal $SU(3)$.  There is an energy gap for production of pairs
above the ground state.  Actually there are two gaps: a smaller one
for the octet, and a larger one for the singlet.

The gluon fields produce an octet of spin $1$ bosons.  As previously
mentioned, they acquire a mass by the Meissner-Higgs phenomenon.  We
have already discussed the Nambu-Goldstone bosons, too.

\subsection{modified photon and integer charges}
The notion of `confinement' I advertised earlier, phrased in terms
of mass gaps and derivative interactions, might seem rather
disembodied.  So it is interesting to ask whether and how a more
traditional and intuitive criterion of confinement -- no fractionally
charged excitations -- is satisfied.

Before discussing electromagnetic charge we must identify the unbroken
gauge symmetry, whose gauge boson defines the physical photon in our
dense medium.  The original electromagnetic gauge invariance is
broken, but there is a combination of the original electromagnetic
gauge symmetry and a color transformation which leaves the condensate
invariant.  Specifically, the original photon $\gamma$ couples
according to the matrix
\begin{equation}
\begin{array}{ccc}
{2 \over 3} & 0 & 0 \\
0& -{1\over 3} & 0 \\
0& 0& -{1\over 3}
\end{array}
\end{equation}
in flavor space, with strength $e$.  There is a gluon $G$ which
couples to the matrix
\begin{equation}
\begin{array}{ccc}
-{2 \over 3} & 0 & 0 \\
0& {1\over 3} & 0 \\
0& 0& {1\over 3}
\end{array}
\end{equation}
in color space, with strength $g$.   Then the combination 
\begin{equation}
\label{photon}
\tilde \gamma = {g \gamma + e G \over \sqrt{e^2 + g^2}}
\end{equation}
leaves the `locking' Kronecker deltas in color-flavor space
invariant.  In our medium, it represents the physical photon.  What
happens here is similar to what occurs in the electroweak sector of
the standard model, where both weak isospin and weak hypercharge are
separately broken by the Higgs doublet, but a cunning combination
remains unbroken, and defines electromagnetism.

Now with respect to $\tilde \gamma$ the electron charge is
\begin{equation}
{-eg\over \sqrt {e^2 + g^2}}~,
\end{equation}
deriving of course solely from the $\gamma$ piece of \ref{photon}.
The quarks have one flavor and one color index, so they pick up
contributions from  both pieces.  In each sector we find the
normalized charge unit ${eg\over \sqrt {e^2 + g^2}}$, and it is
multiplied by some choice from among $(2/3, -1/3, -1/3)$ or $(-2/3,
1/3, 1/3)$ respectively.  The total, obviously, can be $\pm 1$ or $0$.
Thus the excitations produced by the quark fields are integrally
charged, in units of the electron charge.  Similarly the gluons have
an upper color and a lower anti-color index, so that one faces similar
choices, and reaches a similar conclusion.  In particular, some of the
gluons have become electrically charged.  The pseudoscalar
Nambu-Goldstone modes have an upper flavor and a lower anti-flavor
index, and yet again the same conclusions follow.  The superfluid
mode, of course, is electrically neutral.

It is fun to consider how a chunk of our color-flavor locked material
would look.  If the quarks were truly massless, then so would be
Nambu-Goldstone bosons (at the level of pure QCD), and one might
expect a rather unusual `bosonic metal', in which low-energy
electromagnetic response is dominated by these modes.  Actually
electromagnetic radiative corrections lift the mass of the charged
Nambu-Goldstone bosons, creating a gap for the charged channel.  The
same effect would be achieved by turning on a common non-zero quark
mass.  Thus the color-flavor locked material forms a transparent
insulator.  Altogether it resembles a diamond, that reflects portions
of incident light waves, but allows finite portions through and out
again!
\subsection{quark-hadron continuity}

The universal features of the color-flavor locked state: confinement,
chiral symmetry breaking down to vector $SU(3)$, and superfluidity,
are just what one would expect, based on standard phenomenological
models and experience with real-world QCD at low density.  Now we see
that the low-lying spectrum likewise bears an uncanny resemblance to
what one finds in the Particle Data Book (or rather what one would
find, in a world of three degenerate quarks).  It is hard to resist
the inference that there is no phase transition separating them.  Thus
there need not be, and presumably is not, a sharp distinction between
the low-density phase, where microscopic calculations are difficult
but the convenient degrees of freedom are ``obviously'' hadrons,
and the asymptotic high-density phase, where weak-coupling (but
non-perturbative) calculations are possible, and the right degrees of
freedom are elementary quarks and gluons plus collective modes
associated with spontaneous symmetry breaking.  We call this
quark-hadron continuity \cite{cont}.  It might seem shocking that a
quark can ``be'' a baryon, but remember that it is immersed in a
sea of diquark condensate, wherein the distinction between one quark
and three is negotiable.
\subsection{remembrance of things past}
An entertaining aspect of the emergent structure is that two beautiful
ideas from the pre-history of QCD, that were bypassed in its later
development, have come very much back to center stage, now with
microscopic validation.  The quark-baryons of the color-flavor locked
phase follow the charge assignments proposed by Han and Nambu
\cite{han}.  And the gluon-vector mesons derive from the Yang-Mills
gauge principle \cite{yang} -- as originally proposed, for rho mesons!

\section{Fully Microscopic Calculation}
A proper discussion of the fully microscopic calculation \cite{son}
\cite{swMicro} \cite{hong} \cite{pisMicro} is necessarily quite
technical, and would be out of place here, but the spirit of the thing
-- and one of the most striking results -- can be conveyed simply.

When retardation or relativistic effects are important a Hamiltonian
treatment is no longer appropriate.  One must pass to Lagrangian and
graphical methods.  (Theoretical challenge: is it possible to
systematize these in a variational approach?)  The gap equation
appears as a self-consistency equation for the assumed condensation,
shown graphically in Figure 2.

With a contact interaction, and throwing away manifestly spurious
ultraviolet divergences, we obtain a gap equation of the type
\begin{equation}
\Delta \propto  g^2 \int d\epsilon {\Delta \over \sqrt {\epsilon^2 +
\Delta^2}}. 
\end{equation}
The phase space transverse to the fermi surface cancels against a
propagator, leaving the integral over the longitudinal distance
$\epsilon$ to the fermi surface.  Note that the integral on the right
diverges at small $\epsilon$, so that as long as the proportionality
constant is positive one will have non-trivial solutions for $\Delta$,
no matter how small is $g$.  Indeed, one finds that for small $g$,
$\Delta \sim e^{-{\rm const}/g^2}$.

If we restore the gluon propagator, we will find a non-trivial angular
integral, which diverges for forward scattering.  That divergence will
be killed, however, if the gluon acquires a mass $\propto g\Delta$
through the Meissner-Higgs mechanism.  Thus we arrive at a gap
equation of the type
\begin{equation}
\Delta \propto g^2 \int d\epsilon {\Delta \over \sqrt {\epsilon^2 +
\Delta^2}} dz { \mu^2 \over \mu^2 + (g\Delta )^2 }. 
\end{equation}
Now one finds $\Delta \sim e^{-{\rm const}/g}$!

A proper discussion of the microscopic gap equation is considerably
more involved than this, but the conclusion that the gap goes
exponentially in the inverse coupling (rather than its square) at weak
coupling still emerges.  It has the amusing consequence, that at
asymptotically high densities the gap becomes arbitrarily large!
This is because asymptotic freedom insures that it is the microscopic
coupling $1/g(\mu)^2$ which vanishes logarithmically, so that
$e^{-{\rm const}/g(\mu ) }$ does not shrink as fast as $1/\mu$.  Since
the ``dimensional analysis'' scale of the gap is set by $\mu$,
its linear growth wins out asymptotically.

\section{More Quarks}

For larger numbers of quarks, the story is qualitatively similar
\cite{cont}.  Color symmetry is broken completely, and there is a gap
in all quark channels, so the weak-coupling treatment is adequate.
Color-flavor locking is so favorable that there seems to be a
periodicity: if the number of quarks is a multiple of three, one finds
condensation into 3$\times$3 blocks, while if it is 4+3k or 5+3k one
finds k color-flavor locking blocks together with special patterns
characteristic of 4 or 5 flavors.

There is an amusing point here.  QCD with a very large number of
massless quarks, say 16, has an infrared fixed point at very weak
coupling \cite{gross} \cite{caswell}.  Thus it should be quasi-free at
zero density, forming a nonabelian Coulomb phase, featuring conformal
symmetry, no confinement, and no chiral symmetry breaking.  To say the
least, it does not much resemble real-world QCD.  There are
indications that this qualitative behavior may persist even for
considerably fewer quarks (the critical number might be as small as 5
or 6).  Nevertheless, at high density, we have discovered, these
many-quark theories all support more-or-less normal-looking `nuclear
matter' -- including confinement and chiral symmetry breaking!

\section{Fewer Quarks}
\subsection{two flavors}
One can perform a similar analysis for two quark flavors
\cite{arw2SC} \cite{rapp}.  A new feature is that the instanton
interaction now involves four rather than six quark legs, so it
remains relevant as one renormalizes toward the fermi surface.  Either
the one-gluon exchange or the instanton interaction, treated in the
spirit above, favors condensation of the form
\begin{equation}
\label{twoCond}
\begin{array}{lll}
~\langle q^{i\alpha}_{La} (p) q^{j\beta}_{Lb} (-p) \rangle ~=~ 
-\langle q^{i\alpha}_{Ra} (p) q^{j\beta}_{Lb} (-p) \rangle ~=~ 
\epsilon^{ij} \kappa(p^2) \epsilon^{\alpha\beta3} \epsilon_{ab}~.
\end{array}
\end{equation}
Formally, \ref{twoCond} is quite closely related to \ref{condensate},
since $\epsilon^{\alpha\beta I} \epsilon_{abI} = 2(\delta^\alpha_a
\delta^\beta_b - \delta^\alpha_b \delta^\beta_a)$.  Their physical
implications, however, are quite different.

To begin with, \ref{twoCond} does not lead to gaps in all quark
channels.  The quarks with color labels 1 and 2 acquire a gap, but
quarks of the third color of quark are left untouched.  Secondly, the
color symmetry is not completely broken.  A residual $SU(2)$, acting
among the first two colors, remains valid.  For these reasons,
perturbation theory about the ground state defined by \ref{twoCond} is
{\it not\/} free of infrared divergences, and we do not have a fully
reliable grip on the physics.

Nevertheless it is plausible that the qualitative features suggested
by \ref{twoCond} are not grossly misleading.  The residual $SU(2)$
presumably produces confined glueballs of large mass, and assuming
this occurs, the residual gapless quarks are weakly coupled.

Assuming for the moment that no further condensation occurs, for
massless quarks we have the symmetry breaking pattern
\begin{equation}
SU(3)^c \times SU(2)_L \times SU(2)_R \times U(1)_B \rightarrow SU(2)^c \times 
SU(2)_L \times SU(2)_R \times {\tilde U}(1)_B
\end{equation}
Here the modified baryon number acts only on the third color of
quarks.  It is a combination of the original baryon number and a color
generator, that are separately broken but when applied together leave
the condensate invariant.  Comparing to the zero-density ground state,
one sees that color symmetry is reduced, chiral symmetry is restored,
and baryon number is modified.  Only the restoration of chiral
symmetry is associated with a legitimate order parameter, and only it
requires a sharp phase transition.

In the real world, with the $u$ and $d$ quarks light but not strictly
massless, there is no rigorous argument that a phase transition is
necessary.  It is (barely) conceivable that one might extend
quark-hadron continuity to this case \cite{sw2+1}.  Due to medium
modifications of baryon number and electromagnetic charge the
third-color $u$ and $d$ quarks have the quantum numbers of nucleons.
The idea that chiral symmetry is effectively restored in nuclear
matter, however, seems problematic quantitatively.  More plausible,
perhaps, is that there is a first-order transition between nuclear
matter and quark matter.  This is suggested by some model calculations
(e.g., \cite{arw2SC} \cite{rapp}), and is the basis for an attractive
interpretation of the MIT bag model, according to which baryons are
droplets wherein chiral symmetry is restored.

\subsection{thresholds and mismatches}
In the real world there are two quarks, $u$ and $d$, whose mass is
much less than $\Lambda_{QCD}$, and one, $s$, whose mass is comparable
to it.  Two simple qualitative effects, that have major implications
for the zero-temperature phase diagram, arise as consequences of this
asymmetric spectrum \cite{sw2+1} \cite{abr2+1}.  They are expected,
whether one analyzes from the quark side or from the hadron side.

The first is that one can expect a threshold, in chemical potential
(or pressure), for the appearance of any strangeness at all in the
ground state.  This will certainly hold true in the limit of large
strange quark mass, and there is considerable evidence for it in the
real world.  This threshold is in addition to the threshold
transitions at lower chemical potentials, from void to nuclear matter,
and (presumably) from nuclear matter to two-flavor quark matter, as
discussed above.

The second is that at equal chemical potential the fermi surfaces of
the different quarks will not match.  This mismatch cuts off the
Cooper instability in mixed channels.  If the nominal gap is large
compared to the mismatch, one can treat the mismatch as a
perturbation.  This will always be valid at asymptotically high
densities, since the mismatch goes as $m_s^2/\mu$, whereas the gap
eventually grows with $\mu$.  If the nominal gap is small compared to
the mismatch, condensation will not occur.

\subsection{assembling the pieces}
With these complications in mind, we can identify three major phases
in the plane of chemical potential and strange quark mass, that
reflect the simple microscopic physics we have surveyed above.  (There
might of course be additional ``minor'' phases -- notably
including normal nuclear matter!)  There is 2-flavor quark matter,
with restoration of chiral symmetry, and zero strangeness.  Then there
is a 2+1-flavor phase, in which the strange and non-strange fermi
surfaces are badly mismatched, and one has independent dynamics for
the corresponding low energy excitations.  Here one expects
strangeness to break spontaneously, by its own fermi surface
instability.  Finally there is the color-flavor locked phase.  For
some first attempts to sketch a global phase diagram, see \cite{sw2+1}
\cite{abr2+1}.

\section{Comments}

The recent progress, while remarkable, mainly concerns the asymptotic
behavior of QCD.  Its extrapolation to practical densities is at
present semi-quantitative at best.  To do real justice to the
potential applications, we need to learn how to do more accurate
analytical and numerical work at moderate densities.

As regards analytical work, we can take heart from some recent
progress on the equation of state at high temperature \cite{andersen}
\cite{blaizot}.  Here there are extensive, interesting numerical
results \cite{boyd}, which indicate that the behavior is quasi-free,
but that there are very significant quantitative corrections to free
quark-gluon plasma results, especially for the pressure.  Thus it is
plausible {\it a priori\/} that some weak-coupling, but
non-perturbative, approach will be workable, and this seems to be
proving out.  The encouraging feature here is that the analytical
techniques used for high temperature appear to be capable of extension
to finite density without great difficulty.

Numerical work at finite density, unfortunately, is plagued by poor
convergence.  This arises because the functional integral is not
positive definite configuration by configuration, so that importance
sampling fails, and one is left looking for a small residual from much
larger canceling quantities.

There are cases in which this problem does not arise.  It does not
arise for two colors \cite{dagatto}.  Although low-density hadronic
matter is quite different in a two-color world than a three-color
world -- the baryons are bosons, so one does not get anything like a
shell structure for nuclei -- I see no reason to expect that the
asymptotic, high-density phases should be markedly different.  It
would be quite interesting to see fermi-surface behavior arising for
two colors at high density (especially, for the ground state
pressure), and even more interesting to see the effect of diquark
condensations.

Another possibility, that I have been discussing with David Kaplan, is
to engineer lattice gauge theories whose low-energy excitations
resemble those of finite density QCD near the fermi surface, but which
are embedded in a theory that is globally particle-hole symmetric, and
so feature a positive-definite functional integral.

Aside from these tough quantitative issues, there are a number of
directions in which the existing work should be expanded and
generalized, that appear to be quite accessible.  There is already a
rich and important theory of the behavior of QCD at non-zero
temperature and zero baryon number density.  We should construct a
unified picture of the phase structure as a function of both
temperature and density; to make it fully illuminating, we should also
allow at least the strange quark mass to vary.  We should allow for
the effect of electromagnetism (after all, this is largely what makes
neutron stars what they are) and of rotation.  We should consider
other possibilities than a common chemical potential for all the
quarks.

As physicists we should not, however, be satisfied with hoarding up
formal, abstract knowledge.  There are concrete experimental
situations and astrophysical objects we must speak to.  Hopefully,
having mastered some of the basic vocabulary and grammar, we will soon
be in a better position to participate in a two-way dialogue with
Nature.


\newpage

\begin{figure}
\begin{center}
\hspace{0.3cm}
\vspace{0.5cm}
\epsfxsize=8cm
\epsffile{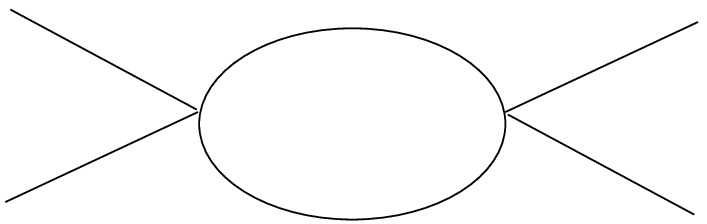}
\end{center}
\vspace*{-1cm}
\caption{Graph contributing to the renormalization of four-fermion
couplings. }
\end{figure}
\newpage


\begin{figure}
\begin{center}
\hspace{0.3cm}
\vspace{0.5cm}
\epsfxsize=12cm
\epsffile{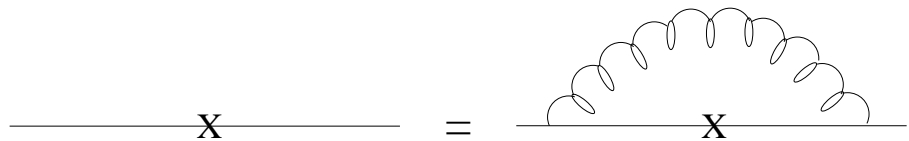}
\end{center}
\vspace*{-1cm}
\caption{Graphical form of the self-consistent equation for the
condensate (gap equation).}
\end{figure}
\newpage


\end{document}